\documentclass[tighten,hyperref={colorlinks=true,linkcolor=blue}]{aastex63}
\usepackage{textcomp}

\usepackage{savesym}

\usepackage[caption=false]{subfig}
\usepackage{longtable}

\usepackage{bm}
\expandafter\ifx\csname package@font\endcsname\relax\else
 \expandafter\expandafter
 \expandafter\usepackage
 \expandafter\expandafter
 \expandafter{\csname package@font\endcsname}
\fi
\savesymbol{tablenum}
\usepackage{siunitx}
\restoresymbol{SIX}{tablenum}
\usepackage{rotating}
\usepackage{tabularx}
\usepackage{graphicx}
\usepackage{wrapfig}
\usepackage{afterpage}

\def\bq{\begin{equation}}
\def\eq{\end{equation}}
\def\bqy{\begin{eqnarray}}
\def\eqy{\end{eqnarray}}

\def\p{\partial}

\def\p{\partial}

\def\R{\mathbb{R}}

 \def\q#1#2{q^{\hspace{1pt}#1}_{\,,\hspace{1pt}#2}}
 \def\a#1#2{a^{\hspace{1pt}#1}_{\,,\hspace{1pt}#2}}
 \def\d#1#2{\delta^{\hspace{1pt}#1}_{\,,\hspace{1pt}#2}}

\shorttitle{Veneras \& Dark Comets}
\shortauthors{Hibberd}

\begin{document}

\title{\large{Study of Venera Spacecraft Trajectories and Wider Implications}}

\correspondingauthor{Adam Hibberd}
\email{adam.hibberd@i4is.org}


\author[0000-0003-1116-576X]{Adam Hibberd}
\affiliation{Initiative for Interstellar Studies (i4is), 27/29 South Lambeth Road London, SW8 1SZ United Kingdom}


\begin{abstract}
Historically, there is no doubt that the early years of the USSR space program put them way ahead of the competition (the USA). Nonetheless, although this was not what the Russians wished to present to the world, the interplanetary campaign, centred around missions to the planet Venus (the \textit{Venera} program) was also beset with difficulties. Many of the early \textit{Venera} probes failed, despite making it to a heliocentric orbit, but naturally the success rate improved with time. The result is that there are now many \textit{Venera} probes in heliocentric orbits, either completely intact, or the main bus after a successful deployment of the lander; together with the associated Blok-L upper stages. This paper is a response to some previous quite contentious research proposing that a certain member of a new class of objects, designated $2005\ VL_1$ may in fact be the \textit{Venera-2} probe. In this paper we look into the invariance of the Earth Tisserand parameter in an attempt to establish if there are indeed any members of this class which could be \textit{Venera} probes. It is found, with extremely small probability, that compared to a sample of randomly chosen NEOs, members of the class of Dark Comet have an Earth Tisserand unusually close to 3, a property shared by the \textit{Venera} missions. Furthermore there are particular associations of 3 Dark Comets with 3 of these probes, the most significant being $2010\ VL_{65}$ with the \textit{Venera-12} mission.
\end{abstract}



\section{Introduction}
\label{sec1}
The discovery of a new class of body, the 'Dark Comet' \citep{2023LPICo2851.2036S,Seligman_2024}, has generated significant debate in the scientific community. These are Near-Earth Objects (NEOs) which have shown signs from orbital analysis, of a significant non-gravitational acceleration (especially perpendicular to the orbital plane, $A_3$) yet exhibit no visible evidence of a coma or tail indicative of outgassing. Their existence would to some extent normalize one of the unusual characteristics of the first interstellar object to be discovered, 1I/'Oumuamua \citep{Flekky2019,Seligman2020,Jackson2021,Desch2021,Bialy2018,Raymond2018}, which also experienced a significant non-gravitational acceleration (radially), with no sign of cometary outgassing \citep{Micheli2018}.\\

This paper explores the plausibility of the attribution of this new class, in the wake of research undertaken by \cite{loeb2025darkcomet2005vl1} indicating one of the members of this new class, designated $2005\ VL_1$, could in fact be the Soviet \textit{Venera-2} probe launched to Venus in 1965.\\

How likely is this to be the case? Is it possible to find supporting evidence to corroborate this claim, or alternatively is there significant counter-evidence to the contrary? This paper looks into the orbital characteristics of 6 of the Dark Comets discovered by \cite{2023LPICo2851.2036S} and attempts to clarify the situation, with further analysis conducted into the orbits of the \textit{Venera} probes as a comparison.

\section{Investigation}
\label{sec2}
\subsection{Overview}
\label{sec2sub1}
The investigations undertaken herein exploit the interplanetary trajectory design tool known as OITS (Optimum Interplanetary Trajectory Software), \cite{OITS_info,AH2}, which incorporates two Non-Linear Problem (NLP) solver options, NOMAD \citep{LeDigabel2011} or MIDACO \citep{Schlueter_et_al_2009,Schlueter_Gerdts_2010,Schlueter_et_al_2013}. This software has been extensively used and demonstrated successfully from applications such as Project Lyra - the feasibility study into spacecraft missions to the first interstellar object to be discovered, 1I/'Oumuamua \citep{HPE19,HEL22,HHE20,HH21,HPH21,AH23,HA23} - to a study into a light sail precursor mission to Enceladus or Europa \citep{LINGAM2024251}. \\

\begin{figure}[hbt!]
\centering
\includegraphics[width=1.0\textwidth]{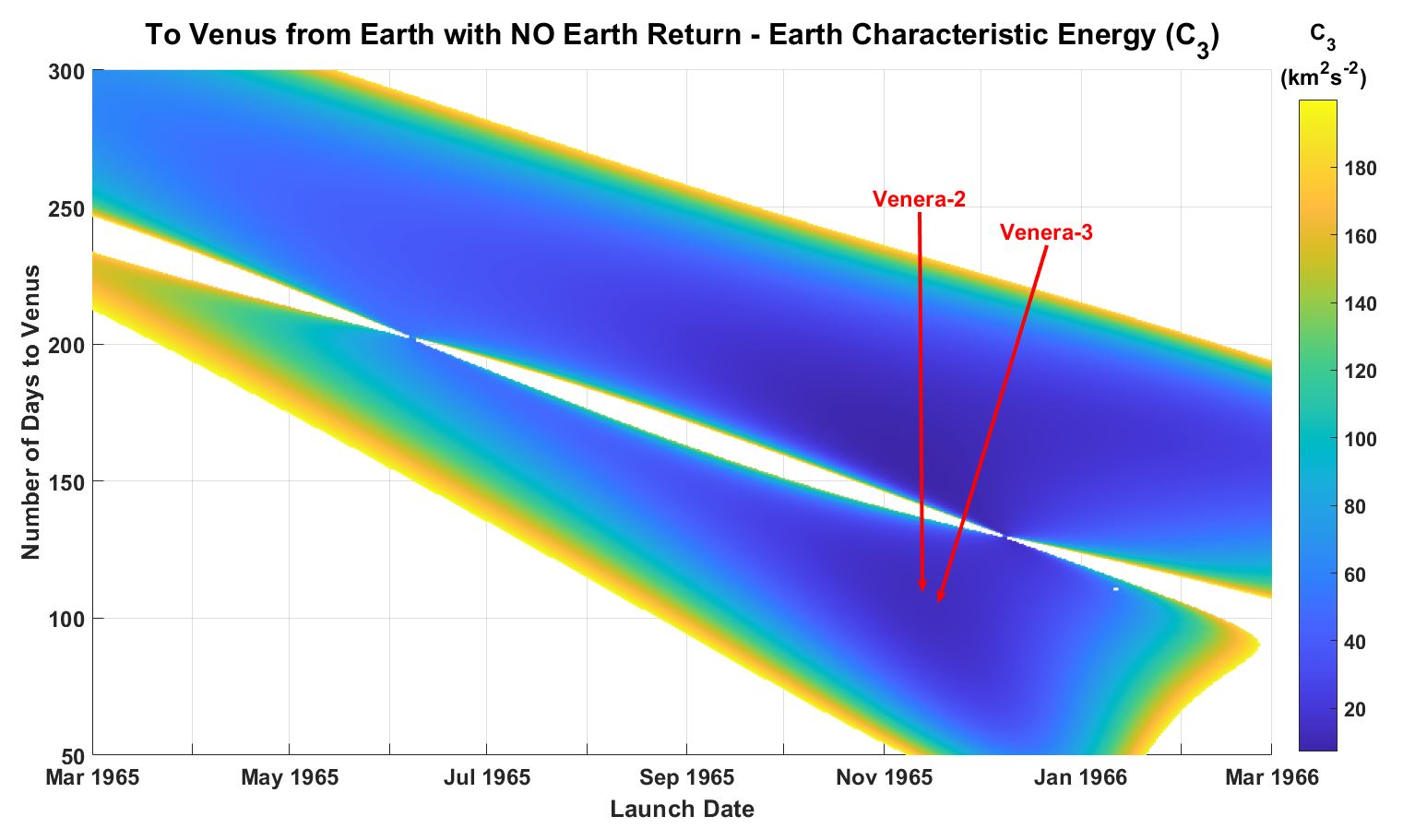}
\caption{Pork chop plot of flyby missions to Venus, with colours representing \textit{Characteristic Energy} ($C_3$) needed by the launch vehicle (or for \textit{Venera-2 \& 3} the Blok-L upper stage) to reach Venus given the launch date (x-axis) and flight duration (y-axis).}
\label{fig:C3_NR}
\end{figure}

We begin by examining a typical 'Pork Chop' plot for flyby missions to Venus assuming NO Earth return, in late 1965, around the time of the \textit{Venera-2} and \textit{Venera-3} missions. This 'Pork Chop' is otherwise known as the colour contour map of required launch vehicle \textit{Characteristic Energy}, $C_3$ (in \si{km^2.s^{-2}}), over the time range in question, refer to Figure \ref{fig:C3_NR}.\\

Observe that for a launch around November 12$^{th}$ (the \textit{Venera-2} launch date) there are two possible regions of optimality (i.e. minima of $C_3$), at a flight time of  $\sim{110}$ days and then again another at $\sim{160}$ days. This is a well understood feature of direct missions to Venus and they are classified as \textit{Type I} and \textit{Type II} trajectories respectively. The latter option represents the superior performance scenario. However for the early period of planetary exploration, the tendency was that the longer the duration of any mission, the more likely it was to fail, thus the Russians favoured the shorter Type I trajectories, despite being significantly suboptimal.\\

To quantify the level of suboptimality, we find that OITS gives the Earth Hyperbolic Excess (V$_{\infty}$, where $C_3 =V_{\infty}^2$) as $\sim{3.65}$ \si{km.s^{-1}}, for a short duration mission, with the arrival at Venus after 110 days. On the contrary, for a more extended duration of 154 days we get the overall optimal V$_{\infty} \sim{2.70}$ \si{km.s^{-1}}. To determine the corresponding extra payload mass to Venus, had the mission planners chosen the lower V$_{\infty}$ route, we examine the exact nature by which the \textit{Venera-2} probe was sent to Venus.\\

According to \cite{Venera2_Bus}, the probe was lofted by a SS-6 (Sapwood) rocket, otherwise known as a Molniya-M (R7-A 8K78) launch vehicle, into a parking orbit (205 $\si{km}$ x 315 $\si{km}$), from which a Blok-L upper stage boosted the \textit{Venera-2} craft into its escape trajectory. The key velocities (including $V_{esc}$, the escape velocity at perigee and $V_{per}$, the velocity at perigee) for this escape trajectory obey the relationship:
\begin{equation}\label{Vesc}
V_{per}^2 = V_{\infty}^2+V_{esc}^2     
\end{equation}
Thus we find that for the shorter mission, $V_{per} = 11.6\ \si{km.s^{-1}}$, whereas for the longer mission,  $V_{per} = 11.3\ \si{km.s^{-1}}$, representing a $\Delta V$ saving of $\sim{0.3}\ \si{km.s^{-1}}$. This corresponds to a Blok-L extra payload capability of $\sim{250}\ \si{kg}$. The \textit{Venera-2} probe had a mass of 963 $\si{kg}$, so had the Russians chosen the optimal route, this would have allowed a spacecraft mass of $\sim{1200} \si{kg}$.

\begin{figure}[hbt!]
\centering
\includegraphics[width=1.0\textwidth]{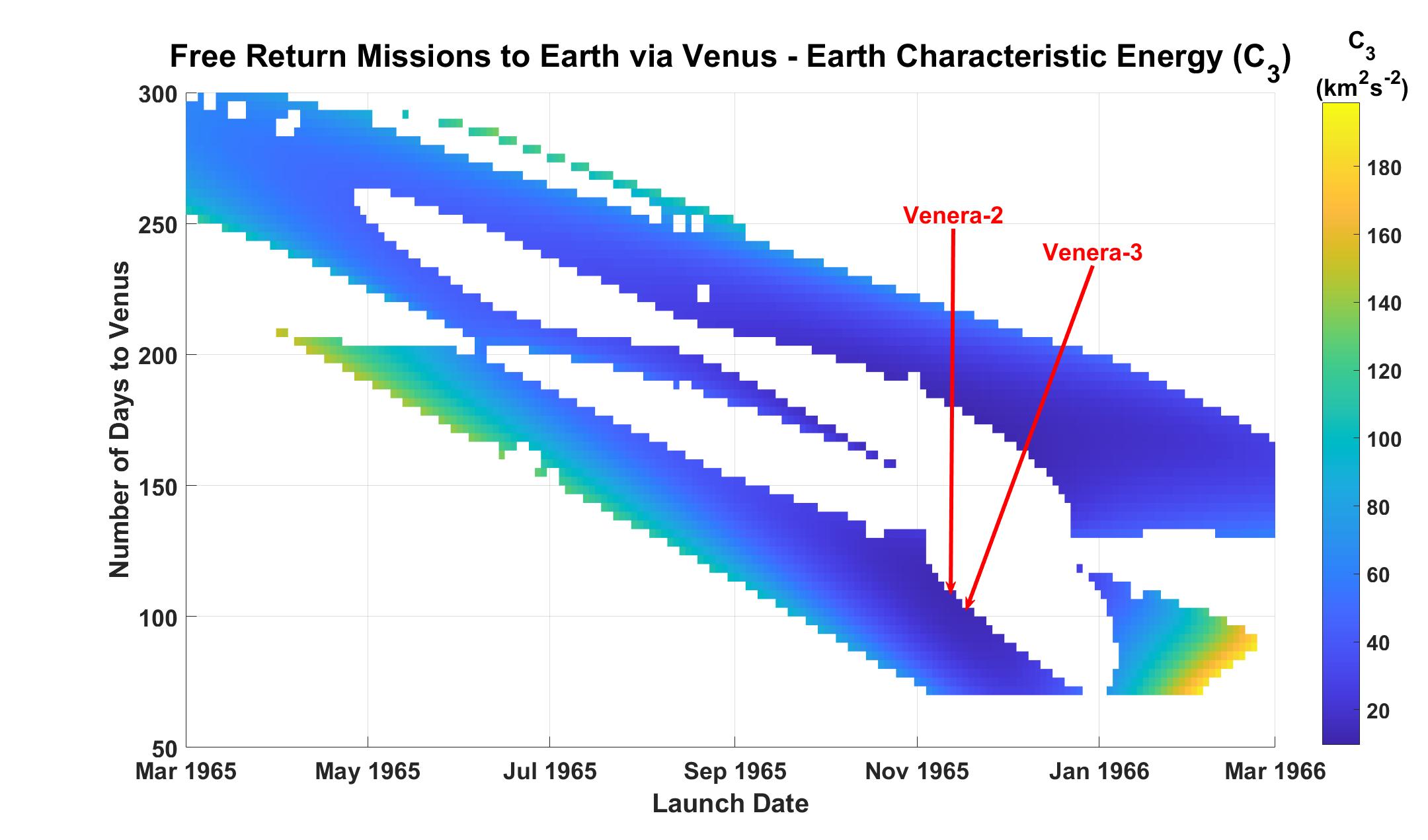}
\caption{Pork chop plot of flyby missions to Venus WITH A SUBSEQUENT RETURN TO EARTH and no thrust required at Venus}
\label{fig:C3_R}
\end{figure}


\subsection{Return to Earth}
\label{sec2sub2}
We now consider the practicalities of a return to Earth from an astrodynamical point-of-view, taking the \textit{Venera-2} and \textit{Venera-3} missions as cases in point.\\

Refer to Figure \ref{fig:C3_R} which shows the same $C_3$ values indicated in Figure \ref{fig:C3_NR}, but removes all Venus flight times that canNOT result in the probe directly returning to Earth because they MUST ENTAIL some propulsion at pericythe.\\

We find that large swathes of landscape occupying the Type II region of Figure \ref{fig:C3_NR}, are now absent because the return trip is infeasible. This demonstrates that the Type I trajectory, although suboptimal from a purely flyby perspective, turns out to be optimal for a return trip.\\

\subsection{The Venera Probes}
\label{sec2sub3}
 
\cite{1985JBIS...38...74C} provides a table (refer Table 1 in that paper) of all the \textit{Venera} probes, in particular their launch and arrival dates. These two items of data are entirely sufficient to determine the interplanetary heliocentric orbital transfer for each probe, ignoring Deep Space Manoeuvres (DSM), and OITS provides an easy facility whereby this calculation can be performed. Refer to Table \ref{table:TAB1} for the data in question, enabling comparison of the positions obtained by Clark with those obtained herein (by Hibberd). This table also provides the resulting orbital elements for each probe determined by OITS using the Universal Variable method of solution \citep{battin1999introduction}.
   
\begin{sidewaystable*}
\hspace{-6.0cm}
\tiny
\renewcommand{\arraystretch}{1.9}
\begin{tabularx}{\textwidth}{|c|c|c|cc|cc|cc|cc|ccccccc|}
\hline
\multicolumn{1}{|l|}{\textbf{}} &
  \multicolumn{1}{l|}{\textbf{}} &
  \multicolumn{1}{l|}{\textbf{}} &
  \multicolumn{2}{c|}{\textbf{Clark}} &
  \multicolumn{2}{c|}{\textbf{Hibberd}} &
  \multicolumn{2}{c|}{\textbf{Clark}} &
  \multicolumn{2}{c|}{\textbf{Hibberd}} &
  \multicolumn{7}{c|}{\textbf{Hibberd}} \\ \hline
\multicolumn{1}{|l|}{\textbf{Spacecraft}} &
  \multicolumn{1}{l|}{\textbf{Launch Date}} &
  \multicolumn{1}{l|}{\textbf{Arrival Date}} &
  \multicolumn{2}{c|}{\textbf{Earth}} &
  \multicolumn{2}{c|}{\textbf{Earth}} &
  \multicolumn{2}{c|}{\textbf{Venus}} &
  \multicolumn{2}{c|}{\textbf{Venus}} &
  \multicolumn{6}{c|}{\textbf{Orbit}} &
  \multicolumn{1}{l|}{\textbf{V$_\infty$}} \\ \hline
\textbf{} &
  \textbf{} &
  \textbf{} &
  \multicolumn{1}{c|}{\textbf{Lon ($^{\circ}$)}} &
  \textbf{Dist (au)} &
  \multicolumn{1}{c|}{\textbf{Lon ($^{\circ}$)}} &
  \textbf{Dist (au)} &
  \multicolumn{1}{c|}{\textbf{Lon ($^{\circ}$)}} &
  \textbf{Dist (au)} &
  \multicolumn{1}{c|}{\textbf{Lon ($^{\circ}$)}} &
  \textbf{Dist (au)} &
  \multicolumn{1}{c|}{\textbf{a (au)}} &
  \multicolumn{1}{c|}{\textbf{q (au)}} &
  \multicolumn{1}{c|}{\textbf{e}} &
  \multicolumn{1}{c|}{\textbf{i ($^{\circ}$)}} &
  \multicolumn{1}{c|}{\textbf{$\Omega$  ($^{\circ}$)}} &
  \multicolumn{1}{c|}{\textbf{$\omega$ ($^{\circ}$)}} &
  \textbf{(km/s)} \\ \hline
Venera-1 &
  12/02/1961 &
  20/05/1961 &
  \multicolumn{1}{c|}{143.0} &
  0.987 &
  \multicolumn{1}{c|}{143.595} &
  0.987 &
  \multicolumn{1}{c|}{263.1} &
  0.727 &
  \multicolumn{1}{c|}{263.797} &
  0.727 &
  \multicolumn{1}{c|}{0.8584} &
  \multicolumn{1}{c|}{0.7082} &
  \multicolumn{1}{c|}{0.175} &
  \multicolumn{1}{c|}{0.47} &
  \multicolumn{1}{c|}{36.08} &
  \multicolumn{1}{c|}{333.72} &
  3.8784 \\ \hline
Zond 1 &
  02/04/1964 &
  19/07/1964 &
  \multicolumn{1}{c|}{192.2} &
  1.000 &
  \multicolumn{1}{c|}{192.790} &
  1.000 &
  \multicolumn{1}{c|}{314.5} &
  0.728 &
  \multicolumn{1}{c|}{315.099} &
  0.728 &
  \multicolumn{1}{c|}{0.8365} &
  \multicolumn{1}{c|}{0.67418} &
  \multicolumn{1}{c|}{0.194} &
  \multicolumn{1}{c|}{3.42} &
  \multicolumn{1}{c|}{12.77} &
  \multicolumn{1}{c|}{359.03} &
  3.5329 \\ \hline
Venera-2 &
  12/11/1965 &
  27/02/1966 &
  \multicolumn{1}{c|}{49.2} &
  0.990 &
  \multicolumn{1}{c|}{50.136} &
  0.990 &
  \multicolumn{1}{c|}{177.2} &
  0.720 &
  \multicolumn{1}{c|}{178.016} &
  0.720 &
  \multicolumn{1}{c|}{0.8421} &
  \multicolumn{1}{c|}{0.6914} &
  \multicolumn{1}{c|}{0.179} &
  \multicolumn{1}{c|}{4.22} &
  \multicolumn{1}{c|}{36.08} &
  \multicolumn{1}{c|}{333.72} &
  3.6503 \\ \hline
Venera-3 &
  16/11/1965 &
  01/03/1966 &
  \multicolumn{1}{c|}{53.3} &
  0.989 &
  \multicolumn{1}{c|}{53.907} &
  0.989 &
  \multicolumn{1}{c|}{180.4} &
  0.720 &
  \multicolumn{1}{c|}{181.061} &
  0.720 &
  \multicolumn{1}{c|}{0.8454} &
  \multicolumn{1}{c|}{0.6958} &
  \multicolumn{1}{c|}{0.177} &
  \multicolumn{1}{c|}{4.13} &
  \multicolumn{1}{c|}{53.85} &
  \multicolumn{1}{c|}{166.33} &
  3.667 \\ \hline
Venera-4 &
  12/06/1967 &
  18/10/1967 &
  \multicolumn{1}{c|}{260.2} &
  1.015 &
  \multicolumn{1}{c|}{260.819} &
  1.015 &
  \multicolumn{1}{c|}{53.9} &
  0.722 &
  \multicolumn{1}{c|}{54.593} &
  0.722 &
  \multicolumn{1}{c|}{0.8666} &
  \multicolumn{1}{c|}{0.7167} &
  \multicolumn{1}{c|}{0.173} &
  \multicolumn{1}{c|}{2.91} &
  \multicolumn{1}{c|}{80.73} &
  \multicolumn{1}{c|}{352.31} &
  2.972 \\ \hline
Venera-5 &
  05/01/1969 &
  16/05/1969 &
  \multicolumn{1}{c|}{104.3} &
  0.983 &
  \multicolumn{1}{c|}{104.918} &
  0.983 &
  \multicolumn{1}{c|}{258.1} &
  0.726 &
  \multicolumn{1}{c|}{258.763} &
  0.726 &
  \multicolumn{1}{c|}{0.8471} &
  \multicolumn{1}{c|}{0.7099} &
  \multicolumn{1}{c|}{0.162} &
  \multicolumn{1}{c|}{0.26} &
  \multicolumn{1}{c|}{74.24} &
  \multicolumn{1}{c|}{6.05} &
  2.845 \\ \hline
Venera-6 &
  10/01/1969 &
  17/05/1969 &
  \multicolumn{1}{c|}{109.6} &
  0.983 &
  \multicolumn{1}{c|}{110.014} &
  0.983 &
  \multicolumn{1}{c|}{259.7} &
  0.726 &
  \multicolumn{1}{c|}{260.349} &
  0.726 &
  \multicolumn{1}{c|}{0.8466} &
  \multicolumn{1}{c|}{0.7103} &
  \multicolumn{1}{c|}{0.161} &
  \multicolumn{1}{c|}{0.42} &
  \multicolumn{1}{c|}{69.48} &
  \multicolumn{1}{c|}{2.67} &
  2.7924 \\ \hline
Venera-7 &
  17/08/1970 &
  15/12/1970 &
  \multicolumn{1}{c|}{323.5} &
  1.012 &
  \multicolumn{1}{c|}{324.095} &
  1.012 &
  \multicolumn{1}{c|}{103.2} &
  0.719 &
  \multicolumn{1}{c|}{103.775} &
  0.719 &
  \multicolumn{1}{c|}{0.8544} &
  \multicolumn{1}{c|}{0.6963} &
  \multicolumn{1}{c|}{0.185} &
  \multicolumn{1}{c|}{2.38} &
  \multicolumn{1}{c|}{35.86} &
  \multicolumn{1}{c|}{176.67} &
  2.931 \\ \hline
Venera-8 &
  17/03/1972 &
  22/07/1972 &
  \multicolumn{1}{c|}{186.2} &
  0.998 &
  \multicolumn{1}{c|}{186.817} &
  0.998 &
  \multicolumn{1}{c|}{320.7} &
  0.728 &
  \multicolumn{1}{c|}{321.253} &
  0.728 &
  \multicolumn{1}{c|}{0.8461} &
  \multicolumn{1}{c|}{0.6938} &
  \multicolumn{1}{c|}{0.18} &
  \multicolumn{1}{c|}{4.29} &
  \multicolumn{1}{c|}{6.81} &
  \multicolumn{1}{c|}{0.7} &
  3.57 \\ \hline
Venera-9 &
  08/06/1975 &
  22/11/1975 &
  \multicolumn{1}{c|}{256.3} &
  1.015 &
  \multicolumn{1}{c|}{256.953} &
  1.015 &
  \multicolumn{1}{c|}{61.8} &
  0.721 &
  \multicolumn{1}{c|}{62.449} &
  0.722 &
  \multicolumn{1}{c|}{0.8680} &
  \multicolumn{1}{c|}{0.7204} &
  \multicolumn{1}{c|}{0.17} &
  \multicolumn{1}{c|}{3.36} &
  \multicolumn{1}{c|}{76.9} &
  \multicolumn{1}{c|}{354.02} &
  3.0166 \\ \hline
Venera-10 &
  14/06/1975 &
  25/10/1975 &
  \multicolumn{1}{c|}{262.1} &
  1.016 &
  \multicolumn{1}{c|}{262.689} &
  1.016 &
  \multicolumn{1}{c|}{66.7} &
  0.721 &
  \multicolumn{1}{c|}{67.274} &
  0.721 &
  \multicolumn{1}{c|}{0.8707} &
  \multicolumn{1}{c|}{0.7209} &
  \multicolumn{1}{c|}{0.172} &
  \multicolumn{1}{c|}{2.11} &
  \multicolumn{1}{c|}{82.6} &
  \multicolumn{1}{c|}{348.84} &
  2.8769 \\ \hline
Venera-11 &
  09/09/1978 &
  25/12/1978 &
  \multicolumn{1}{c|}{345.6} &
  1.007 &
  \multicolumn{1}{c|}{346.271} &
  1.007 &
  \multicolumn{1}{c|}{120.8} &
  0.718 &
  \multicolumn{1}{c|}{121.436} &
  0.791 &
  \multicolumn{1}{c|}{0.8738} &
  \multicolumn{1}{c|}{0.7139} &
  \multicolumn{1}{c|}{0.183} &
  \multicolumn{1}{c|}{3.39} &
  \multicolumn{1}{c|}{13.72} &
  \multicolumn{1}{c|}{151.58} &
  3.7654 \\ \hline
Venera-12 &
  14/09/1978 &
  21/12/1978 &
  \multicolumn{1}{c|}{350.5} &
  1.006 &
  \multicolumn{1}{c|}{351.135} &
  1.006 &
  \multicolumn{1}{c|}{114.4} &
  0.719 &
  \multicolumn{1}{c|}{114.943} &
  0.719 &
  \multicolumn{1}{c|}{0.8790} &
  \multicolumn{1}{c|}{0.7111} &
  \multicolumn{1}{c|}{0.191} &
  \multicolumn{1}{c|}{2.67} &
  \multicolumn{1}{c|}{8.86} &
  \multicolumn{1}{c|}{146.37} &
  4.0062 \\ \hline
Venera-13 &
  30/10/1981 &
  01/03/1982 &
  \multicolumn{1}{c|}{36.1} &
  0.993 &
  \multicolumn{1}{c|}{36.742} &
  0.993 &
  \multicolumn{1}{c|}{183.3} &
  0.720 &
  \multicolumn{1}{c|}{183.942} &
  0.720 &
  \multicolumn{1}{c|}{0.8514} &
  \multicolumn{1}{c|}{0.7084} &
  \multicolumn{1}{c|}{0.168} &
  \multicolumn{1}{c|}{5.97} &
  \multicolumn{1}{c|}{36.73} &
  \multicolumn{1}{c|}{174.89} &
  4.0291 \\ \hline
Venera-14 &
  04/11/1981 &
  05/03/1982 &
  \multicolumn{1}{c|}{41.1} &
  0.992 &
  \multicolumn{1}{c|}{41.746} &
  0.992 &
  \multicolumn{1}{c|}{189.7} &
  0.721 &
  \multicolumn{1}{c|}{190.408} &
  0.721 &
  \multicolumn{1}{c|}{0.8536} &
  \multicolumn{1}{c|}{0.7136} &
  \multicolumn{1}{c|}{0.164} &
  \multicolumn{1}{c|}{5.97} &
  \multicolumn{1}{c|}{41.73} &
  \multicolumn{1}{c|}{170.19} &
  4.0207 \\ \hline
Venera-15 &
  02/06/1983 &
  10/10/1983 &
  \multicolumn{1}{c|}{250.5} &
  1.014 &
  \multicolumn{1}{c|}{251.157} &
  1.014 &
  \multicolumn{1}{c|}{43.9} &
  0.723 &
  \multicolumn{1}{c|}{46.223} &
  0.723 &
  \multicolumn{1}{c|}{0.8646} &
  \multicolumn{1}{c|}{0.715} &
  \multicolumn{1}{c|}{0.173} &
  \multicolumn{1}{c|}{4.09} &
  \multicolumn{1}{c|}{71.13} &
  \multicolumn{1}{c|}{357.24} &
  3.2247 \\ \hline
Venera-16 &
  07/06/1983 &
  14/10/1983 &
  \multicolumn{1}{c|}{255.3} &
  1.015 &
  \multicolumn{1}{c|}{255.944} &
  1.015 &
  \multicolumn{1}{c|}{50.3} &
  0.722 &
  \multicolumn{1}{c|}{51.033} &
  0.722 &
  \multicolumn{1}{c|}{0.8665} &
  \multicolumn{1}{c|}{0.7175} &
  \multicolumn{1}{c|}{0.172} &
  \multicolumn{1}{c|}{3.5} &
  \multicolumn{1}{c|}{75.91} &
  \multicolumn{1}{c|}{352.72} &
  3.1166 \\ \hline
 &
   &
   &
  \multicolumn{1}{c|}{} &
   &
  \multicolumn{1}{c|}{} &
   &
  \multicolumn{1}{c|}{} &
   &
  \multicolumn{1}{c|}{} &
   &
  \multicolumn{1}{c|}{} &
  \multicolumn{1}{c|}{} &
  \multicolumn{1}{c|}{} &
  \multicolumn{1}{c|}{} &
  \multicolumn{1}{c|}{} &
  \multicolumn{1}{c|}{} &
   \\ \hline
$2005\ VL_1$ &
   &
   &
  \multicolumn{1}{c|}{} &
   &
  \multicolumn{1}{c|}{} &
   &
  \multicolumn{1}{c|}{} &
   &
  \multicolumn{1}{c|}{} &
   &
  \multicolumn{1}{c|}{0.891168} &
  \multicolumn{1}{c|}{0.690983} &
  \multicolumn{1}{c|}{0.224633} &
  \multicolumn{1}{c|}{0.235735} &
  \multicolumn{1}{c|}{} &
  \multicolumn{1}{c|}{} & 
   \\ \hline
\end{tabularx}

\caption{Venera Missions\label{table:TAB1}}
\end{sidewaystable*}

\subsection{Dark Comets and Tisserands}
\label{sec2sub4}

 As expounded in \cite{2023LPICo2851.2036S,Seligman_2024}, over the last few years there has been the discovery of a new class of celestial object. Objects belonging to this proposed new class, the 'Dark Comets', have shown no sign of outgassing, as would be evidenced through a coma or tail, yet they have significant non-gravitational accelerations, particularly so in the $A_3$ component perpendicular to the orbital plane, refer to Table 1 of  \cite{2023LPICo2851.2036S}.\\
 
 Furthermore a study conducted by \cite{loeb2025darkcomet2005vl1} has 
 suggested that one of these objects, $2005\ VL_1$ is in fact the \textit{Venera-2} probe (listed here on the third row of Table \ref{table:TAB1}), although yet a further paper \cite{McDowell_2025} disputes their conclusion. \\

 The main argument put forward by \cite{McDowell_2025}, is that there is a significant and irreconcilable difference between the orbital inclination of the \textit{Venera-2} probe and the Dark Comet $2005\ VL_1$. Table \ref{TAB3} shows the pertinent orbital parameters of 6 Dark Comets studied by \cite{2023LPICo2851.2036S}.\\

We find that there is indeed a difference in the inclinations of \textit{Venera-2} ($4.22^{\circ}$) and $2005\ VL_1$ ($0.24^{\circ}$). However, the significance of this depends on whether the probe returned to Earth and then was affected by Earth's gravitational field. If so, then this encounter could well have flattened the probe's heliocentric orbital plane. Moreover, as already established in this paper, a return to Earth is well within the realms of feasibility for \textit{Venera} probes, refer to Section  \ref{sec2sub2}.\\

\begin{table}[]
\renewcommand{\arraystretch}{2}
\centering
\hspace{-2.0cm}
\begin{tabular}{|ccccccc|}
\hline
\textbf{Dark Comet} & \textbf{a (au)} & \textbf{e} & \textbf{q (au)} & \textbf{Q (au)} & \textbf{i ($^{\circ}$)} & \textbf{$T_E$} \\ \hline
\textbf{2005 VL$_1$}    & 0.891644 & 0.225386 & 0.690680 & 1.092607 & 0.247217  & 2.961452 \\ \hline
\textbf{2016 NJ$_{33}$} & 1.311729 & 0.208452 & 1.038296 & 1.585162 & 6.640499  & 2.987619 \\ \hline
\textbf{2010 VL$_{65}$} & 1.069275 & 0.143999 & 0.915301 & 1.223249 & 4.713182  & 2.9748536 \\ \hline
\textbf{2010 RF$_{12}$} &  1.060512 & 0.188300 & 0.860817 & 1.260207 & 0.882622  & 2.9654805 \\ \hline
\textbf{2006 RH$_{120}$}& 0.995959 & 0.035748 & 0.960355 & 1.031562 & 0.283106  & 2.9987122 \\ \hline
\textbf{2003 RM}  & 2.921126 & 0.601025 & 1.165457 & 4.676795 & 10.853677 & 3.0254402 \\ \hline
\end{tabular}
\caption{Some Dark Comets and pertinent orbital parameters\label{TAB3}}
\end{table}

There is a function of the heliocentric orbital parameters which is approximately invariant before and after an encounter with a planet. Determination of its value can help establish whether two objects observed separately at different times, with different heliocentric orbits, are in fact the same object whose heliocentric orbital path has been altered by the encounter. This invariant is known as the \textit{Tisserand parameter} \citep{SSDYN,Haranas2023}.\\

The Tisserand with respect to Earth, $T_E$ is calculated as :

\begin{equation}\label{TissE}
T_E =\frac{AU}{a} +2\cos{i}\sqrt{\frac{a}{AU}(1-e^2)},     
\end{equation}

where  AU is the astronomical unit, and a is the semi-major axis, i is the inclination and e the eccentricity of the object in question.\\

\begin{figure}[hbt!]
\centering
\includegraphics[width=0.8\textwidth]{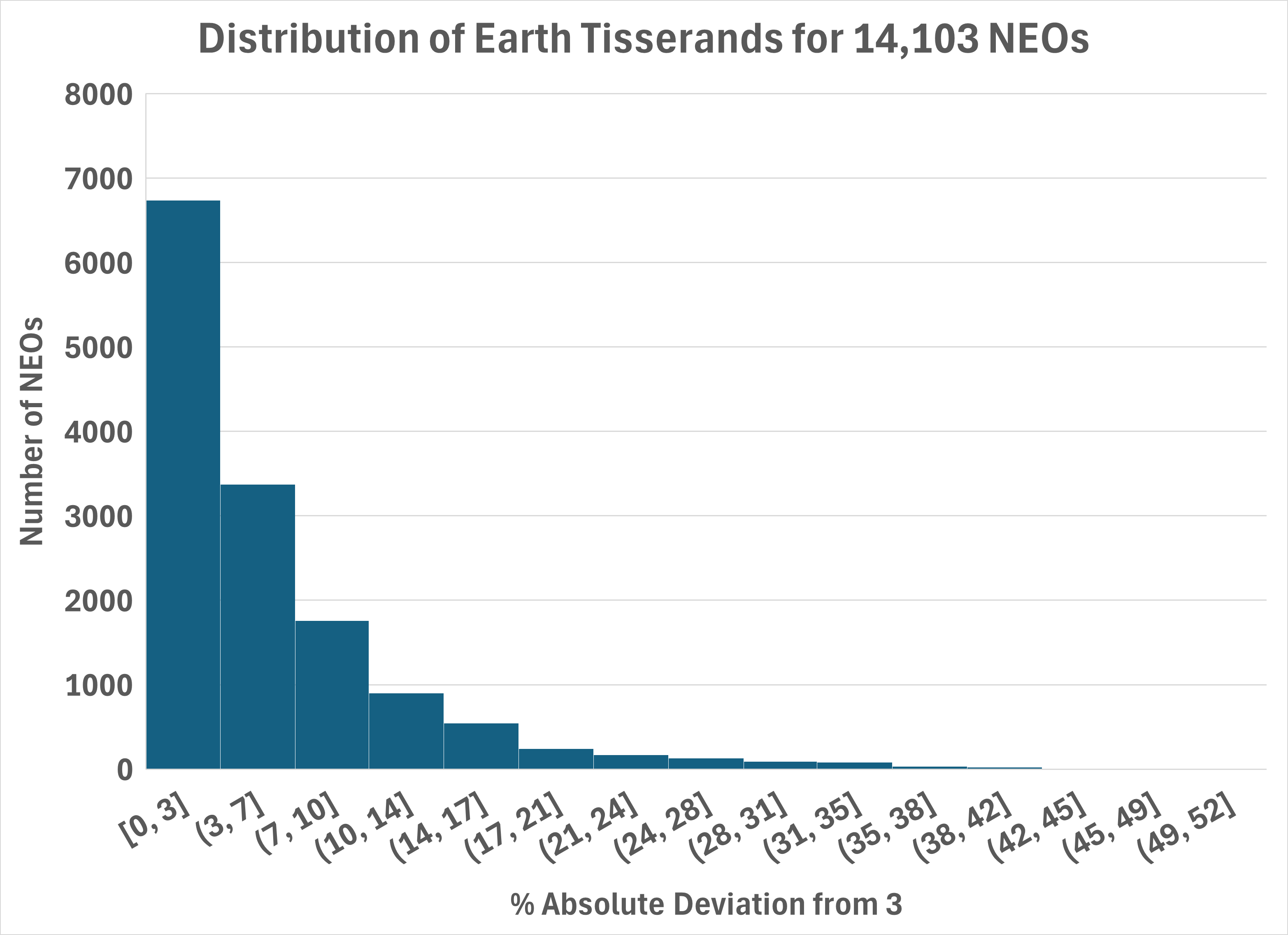}
\caption{Histogram of Earth Tisserands compared to a value of 3, for 14,103 randomly selected NEOs}
\label{fig:Hist2}
\end{figure}

With the intention of establishing a reference sample let us take 14,103 selected Near-Earth Objects (filtered such that 0.5 au $<$ a $<$ 1.5 au and 0.2 au $<$ q $<$ 1.1 au) and determine their range of $T_E$. Refer to the histogram in Figure \ref{fig:Hist2}. Note this is their absolute deviation from a value of 3. We find that there are deviations from this value as much as $\sim{30\%}$ or more.\\
 
\begin{figure}[hbt!]
\centering
\includegraphics[width=0.8\textwidth]{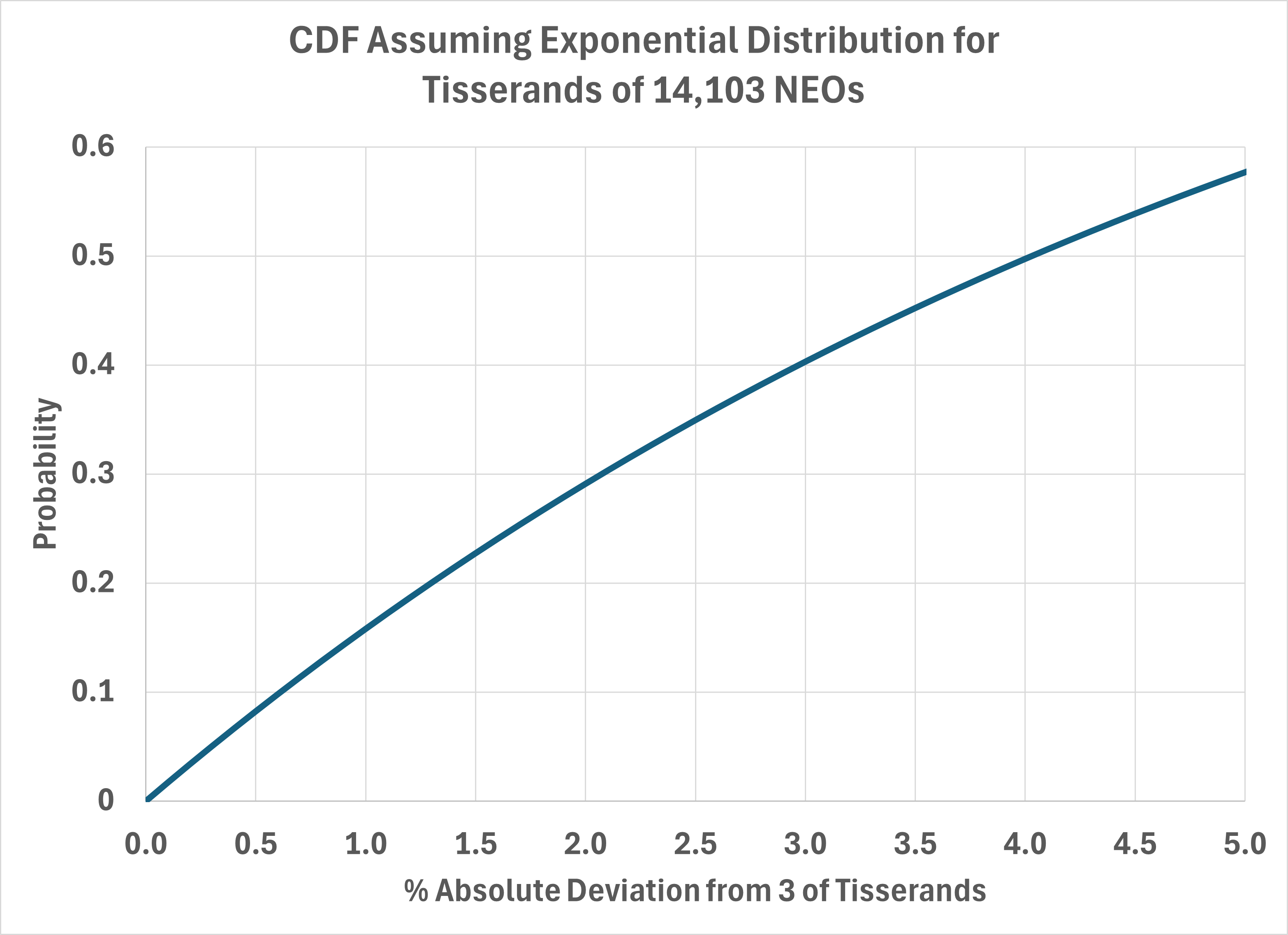}
\caption{CDF of best fit exponentially distributed Tisserands compared to a value of 3, for 14,310 randomly selected NEOs}
\label{fig:Expbf}
\end{figure}

We find in Table \ref{TAB3} that the Tisserands for the 6 Dark Comets are very close to a value of 3, although the significance of this is at the moment unknown since Figure \ref{fig:Hist2} indicates that the most likely $T_E$ for an NEO will indeed be $\sim{3}$. If we assume an Exponential distribution of this data ($\mu = 5.81 \%$) the cumulative distribution function (CDF) is provided in Figure \ref{fig:Expbf}.\\

\begin{figure}[hbt!]
\centering
\includegraphics[width=1.0\textwidth]{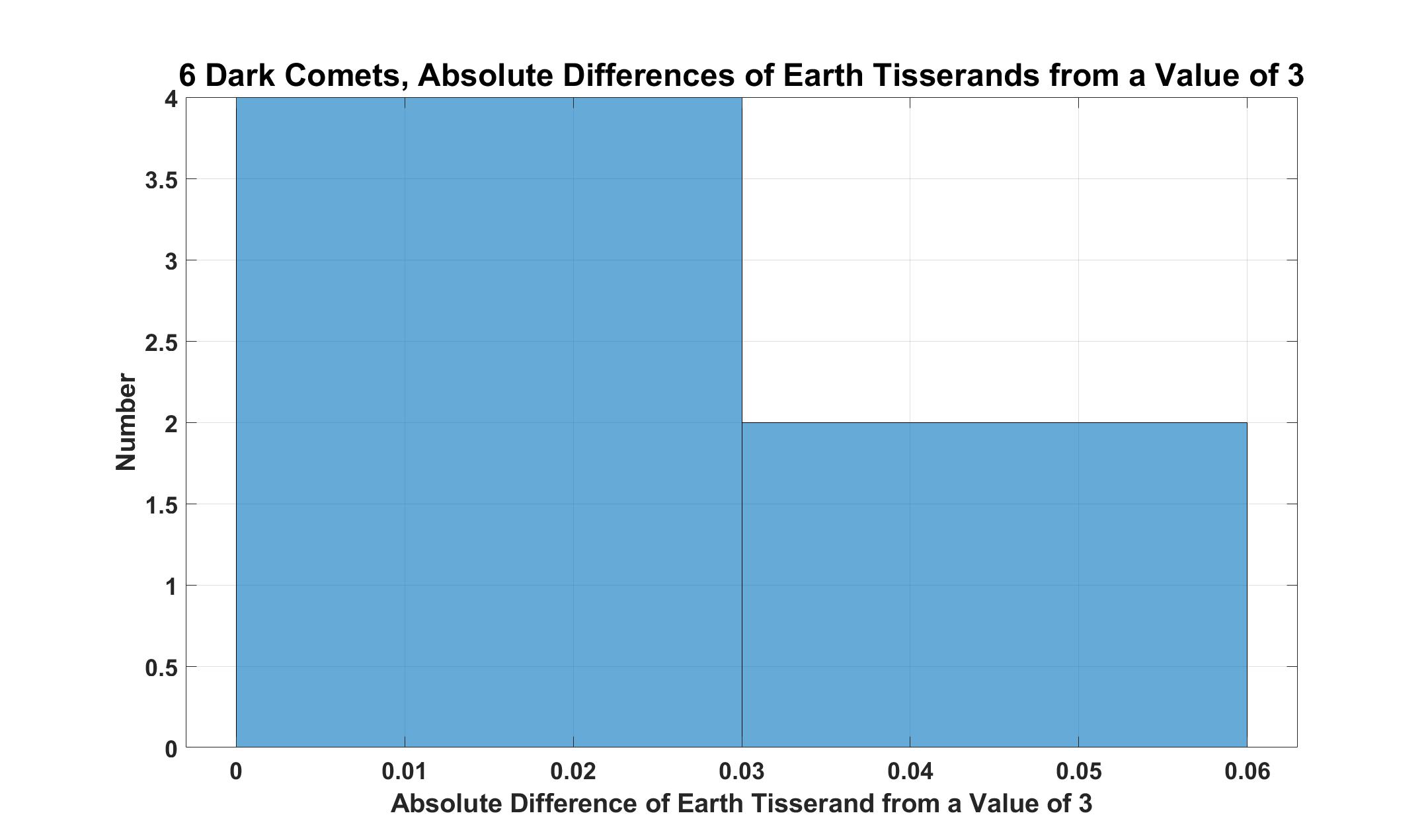}
\caption{Histogram of Earth Tisserands compared to a value of 3, for 6 Dark Comets}
\label{fig:Hist3}
\end{figure}

We now take the Tisserand of all of the 6 Dark Comets and calculate their discrepancy from 3, and we get the histogram in Figure \ref{fig:Hist3}. Observe that the deviations from a value of 3 of these Tisserands are all $<$ 0.06 (equivalent to 2\%). From Figure \ref{fig:Expbf} we find that the probability that a randomly selected NEO has an Earth Tisserand different from 3 by $<$ 2\% is approx. 0.3, in which case the likelihood that 6 randomly chosen NEOs are all this different would be approx. $0.3^6 \sim{0.0007}$. Thus we can conclude that the fact that all 6 Dark Comets have a Tisserand so close to 3 is highly significant, and would be extremely unlikely to happen through pure chance. 


\subsection{Invariance of the Earth Tisserand}
\label{sec2sub5}
We now examine the degree to which the Earth Tisserand, $T_E$ changes as a result of an encounter with Earth in the situation where $T_E \sim{3}$, as for the \textit{Venera} probes. Although in principle approximately invariant, nevertheless before we continue it would be instructive to determine what in reality is the typical level of alteration in $T_E$ measured before and after an Earth encounter. The timescales here are on the order of $\sim{50}$ years (the time elapsed since the launch of the \textit{Veneras}). On top of any theoretical change (it is important to note that the Tisserand is in THEORY only an approximate invariant), this modification to $T_E$ would also be due to orbital perturbations, for the two reasons below:

\begin{figure}[hbt!]
\centering
\includegraphics[width=0.9\textwidth]{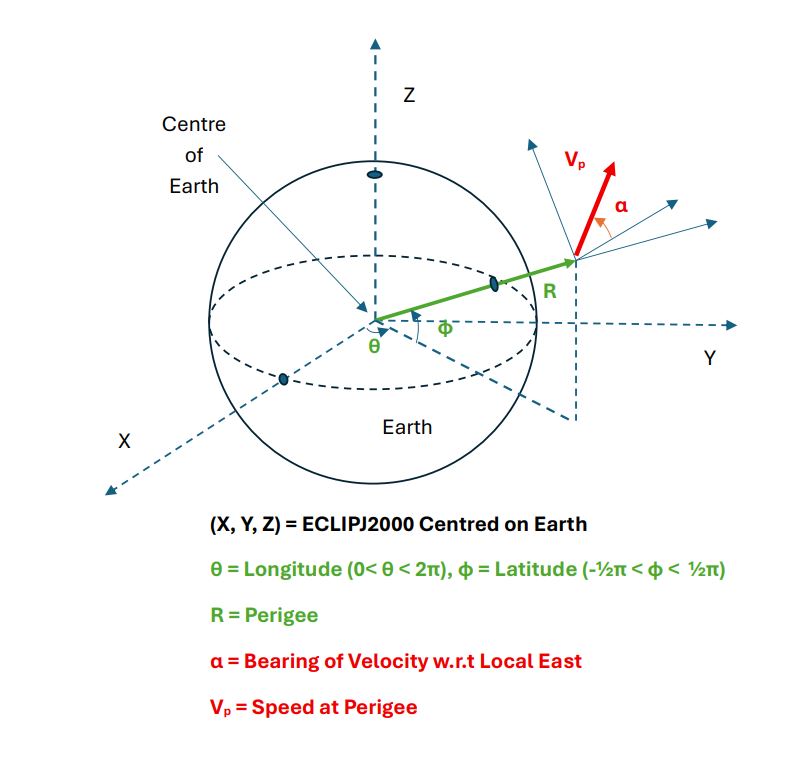}
\caption{Initial set-up of test particle relative to Earth to determine change in Earth Tisserand, $T_E$}
\label{fig:AXES}
\end{figure}

\begin{enumerate}
    \item{the gravitational influence of the 8 planets and our Moon}
    \item{the possible presence of non-gravitational terms, $A_1$, $A_2$, $A_3$ (radial, transverse and out-of-plane respectively) as discovered on the 6 Dark Comets by \cite{2023LPICo2851.2036S}, and whose equations are laid forth in \cite{1973AJ.....78..211M}.}
\end{enumerate}

To this end, the software library known as REBOUND \citep{2012A&A...537A.128R,2015MNRAS.446.1424R} was exploited in conjunction with the NASA JPL NAIF SPICE library \citep{ACTON199665,ACTON20189}. REBOUND is an N-body gravitational integrator, enabling forwards and backwards integration for a system of N-bodies from any point in time. The SPICE library (amongst other things) allows accurate determination of Solar System planetary ephemerides at any time. Thus in conjunction, these libraries can enable forwards and backwards integration from any initial Solar System 'state'.\\

\begin{table}[]
\renewcommand{\arraystretch}{1.6}
\begin{tabular}{|c|c|c|c|}

\hline
\textbf{Parameter} &
  \textbf{Description} &
  \textbf{Range} &
  \textbf{Distribution} \\ \hline
\textbf{R} &
  Perigee Radius &
  \begin{tabular}[c]{@{}c@{}}6378 km + 10,000 km\\ to\\ 6378 km + 1,500,000 km\end{tabular} &
  Uniform \\ \hline
\textbf{$V_p$} &
  Perigee Velocity &
  Determined by $V_{\infty}$ &
  See below \\ \hline
\textbf{$V_{\infty}$} &
  Hyperbolic Excess &
  2.7 $\si{kms^{-1}}$ to 4.1 $\si{kms^{-1}}$ &
  Uniform \\ \hline
\textbf{$\alpha$} &
  Bearing of $V_p$ &
  -$\pi$ to +$\pi$ &
  Uniform \\ \hline
\textbf{$\theta$} &
  \begin{tabular}[c]{@{}c@{}}Ecliptic Longitude\\ of Perigee\end{tabular} &
  0   to   +2$\pi$ &
  Uniform \\ \hline
\textbf{$\phi$} &
  \begin{tabular}[c]{@{}c@{}}Ecliptic Latitude\\ of Perigee\end{tabular} &
  $-\frac{\pi}{2}$ to +$\frac{\pi}{2}$ &
  Uniform \\ \hline
  \textbf{$T_{init}$} &
  \begin{tabular}[c]{@{}c@{}}Time of Start\\ of Simulation \end{tabular} &
  01/01/2000 $\pm{1.0}$ year &
  Uniform \\ \hline
  \textbf{$T_{extrap}$} &
  \begin{tabular}[c]{@{}c@{}}Exrapolation\\ Time\end{tabular} &
  $\pm{25}$ years &
  First forwards then back\\ \hline  
\end{tabular}
\caption{Table of initial conditions for test particle given in Figure \ref{fig:AXES} with their distributions\label{initcond}}
\end{table}
\begin{figure}
     \centering
     \subfloat[All 10,000 objects with Earth Tisserands $\sim{3}$]{
         \includegraphics[width=0.75\linewidth]{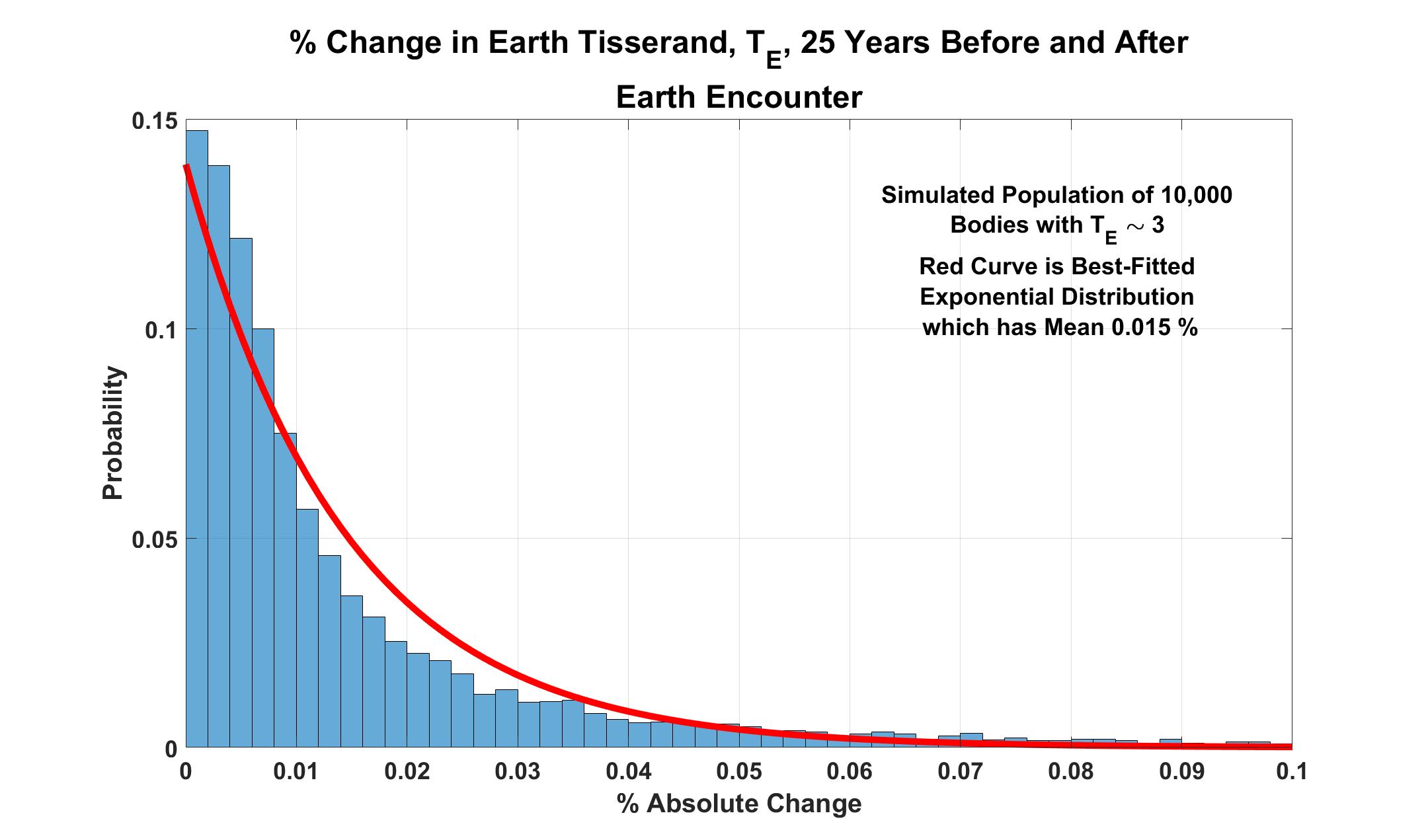}
         \label{fig:1a}
}
\vfill
     \subfloat[Only those objects with with typical \textit{Venera} semi-major axes]{
         \includegraphics[width=0.75\linewidth]{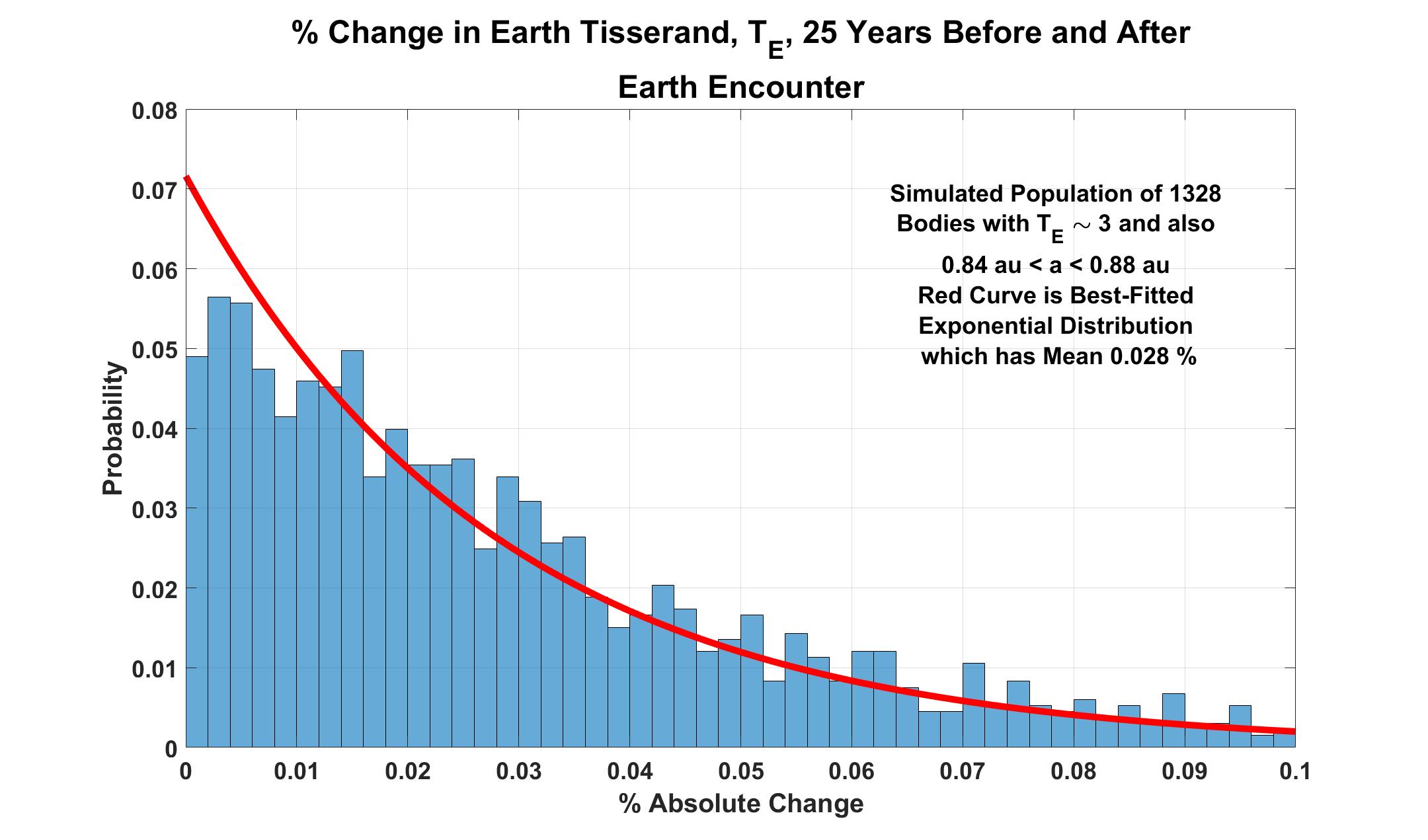}
           \label{fig:1b}
}
     \caption{Absolute Change in Earth Tisserand assuming NO non-gravitational forces (see Section \ref{sec2sub5} item 1) }
     \label{fig:TISS1}
\end{figure}
\begin{figure}
     \centering
     \subfloat[All 10,000 objects with Earth Tisserands $\sim{3}$]{
         \includegraphics[width=0.75\linewidth]{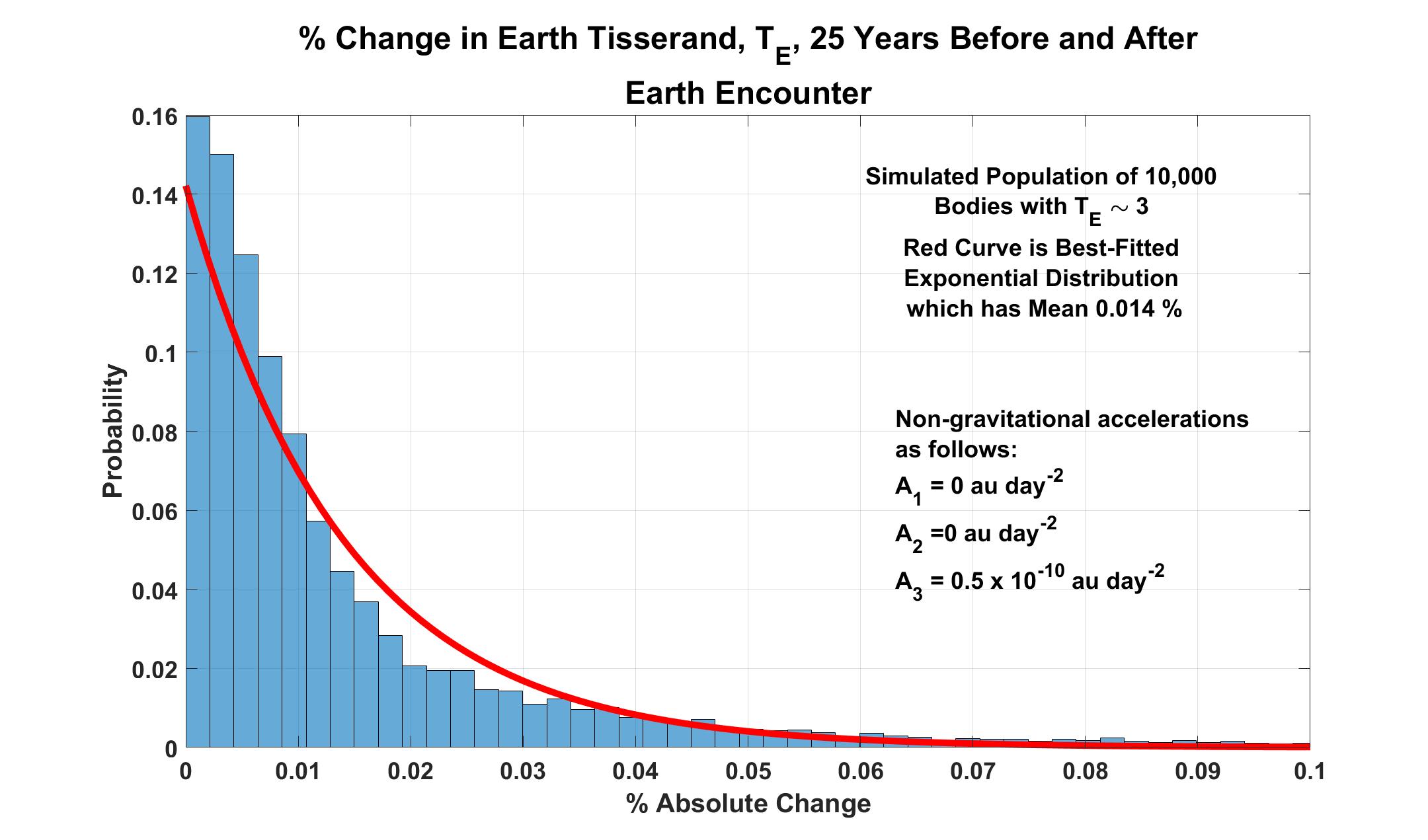}
         \label{fig:2a}
}
\vfill
     \subfloat[Only those objects with typical \textit{Venera} semi-major axes]{
         \includegraphics[width=0.75\linewidth]{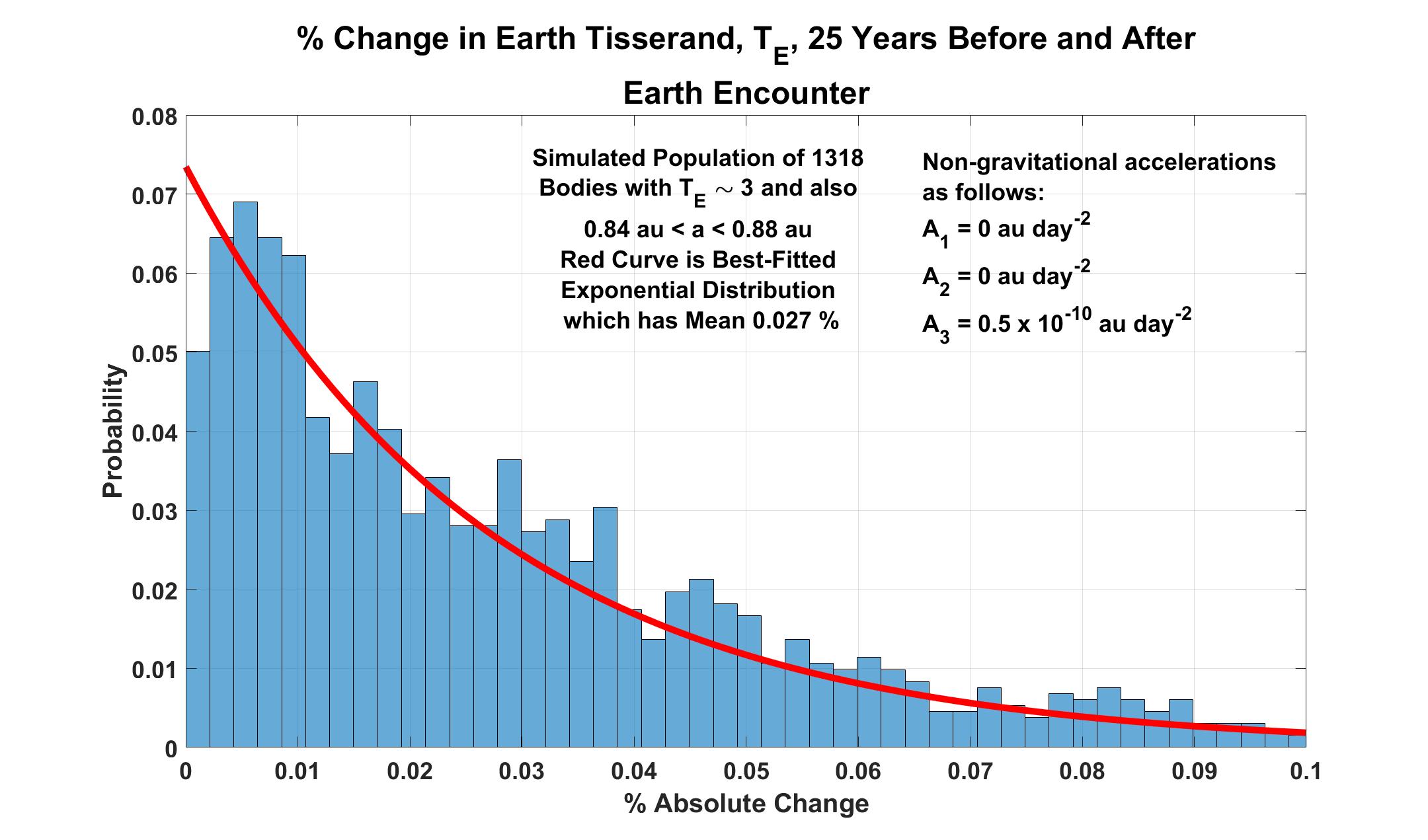}
         \label{fig:2b}
}
     \caption{Absolute Change in Earth Tisserand in the presence of non-gravitational forces (see Section \ref{sec2sub5} item 2)}
     \label{fig:TISS2}
\end{figure}

The introduction of a 'test particle' can be arranged, this being a random body in whose $T_E$ we are interested. It is placed initially near to Earth, at perigee, with a horizontal velocity, $V_p$ whose bearing is given by $\alpha$ (ref. Figure \ref{fig:AXES}). The perigee radius R, longitude, $\theta$ and latitude, $\phi$ (w.r.t the SPICE ECLIPJ2000 reference frame centred on Earth), define the initial location relative to Earth of this test particle. The simulation is then sent forwards in time by 25 years, its Tisserand calculated, and then from the same point, backwards in time for 25 years, whence the change in Tisserand, $\Delta T_E$ is determined. This process is repeated 9999 times, each time with different starting parameters, $\theta$, $\phi$, $R$, $\alpha$, $V_p$, $T_{init}$ amounting to 10,000 test runs, each with a forwards and backwards time-arrow, making 20,000 altogether, with a Monte-Carlo simulation. The distributions of all pertinent parameters for this Monte-Carlo analysis are provided in Table \ref{initcond}.\\

The results of this analysis can be broken down into 4 plots provided in Figures \ref{fig:TISS1} and \ref{fig:TISS2}. Note the choice of parameters $A_1$, $A_2$ and $A_3$ are simply those typical of the Dark Comets in Table 1 of \cite{2023LPICo2851.2036S}, with $A_1 = 0$, $A_2 = 0$ and $A_3 = 0.5\times 10^{-10}$ (units are all $au  \ day^{-2}$) since on-the-whole these Dark Comets only have significant levels of non-gravitational acceleration perpendicular to their orbital planes (thus $A_3 \ne 0$).\\

When we conduct this analysis we find, as perhaps would be expected, that the deviations themselves do not follow a Normal distribution at all closely, because the best-fit Normal tends to underestimate the probability of the change in Tisserand where $\Delta T_E \sim{0}$. An Exponential distribution of the absolute deviations on the other hand provides a substantially closer fit.\\

Referring to Figures \ref{fig:TISS1} and \ref{fig:TISS2}, we see that the mean absolute \% change of the Tisserands, is typically around 0.015\% for the entire population of 10,000 test particles, yet significantly larger, around 0.028\% for the two subsets of the populations where the semi-major axes resemble those of the \textit{Venera} probes.\\ 

\subsection{Comparison of Dark Comets with Venera probes}
\label{sec2sub6}

An important observation is made here in that the Earth Tisserand is significantly NOT invariant to an encounter with Venus where the pericythe altitude (closest approach/periapsis to Venus) was typically around 35,000 \si{km}, for the \textit{Venera} probes \citep{1985JBIS...38...74C}, or to be more accurate, the 'buses' (after the landers had been deployed).\\

Thus any similarity in Earth Tisserand between a \textit{Venera} probe and a Dark Comet would NOT necessarily indicate the object to be a \textit{Venera} probe. However the previously mentioned Blok-L upper stages (refer Section \ref{sec2sub1}) designed for accelerating the \textit{Veneras} from their temporary Earth parking orbit to their interplanetary escape orbits, could well have flown by Venus at much greater distances, and been unaffected, thus this would seem to be a more sensible attribution. Henceforth in order to maintain clarity and still identify each mission individually, we shall assume that by a \textit{Venera-n} probe we mean that it is most likely the Blok-L upper stage for the corresponding mission.\\

\begin{figure}[hbt!]
\centering
\includegraphics[width=0.8\textwidth]{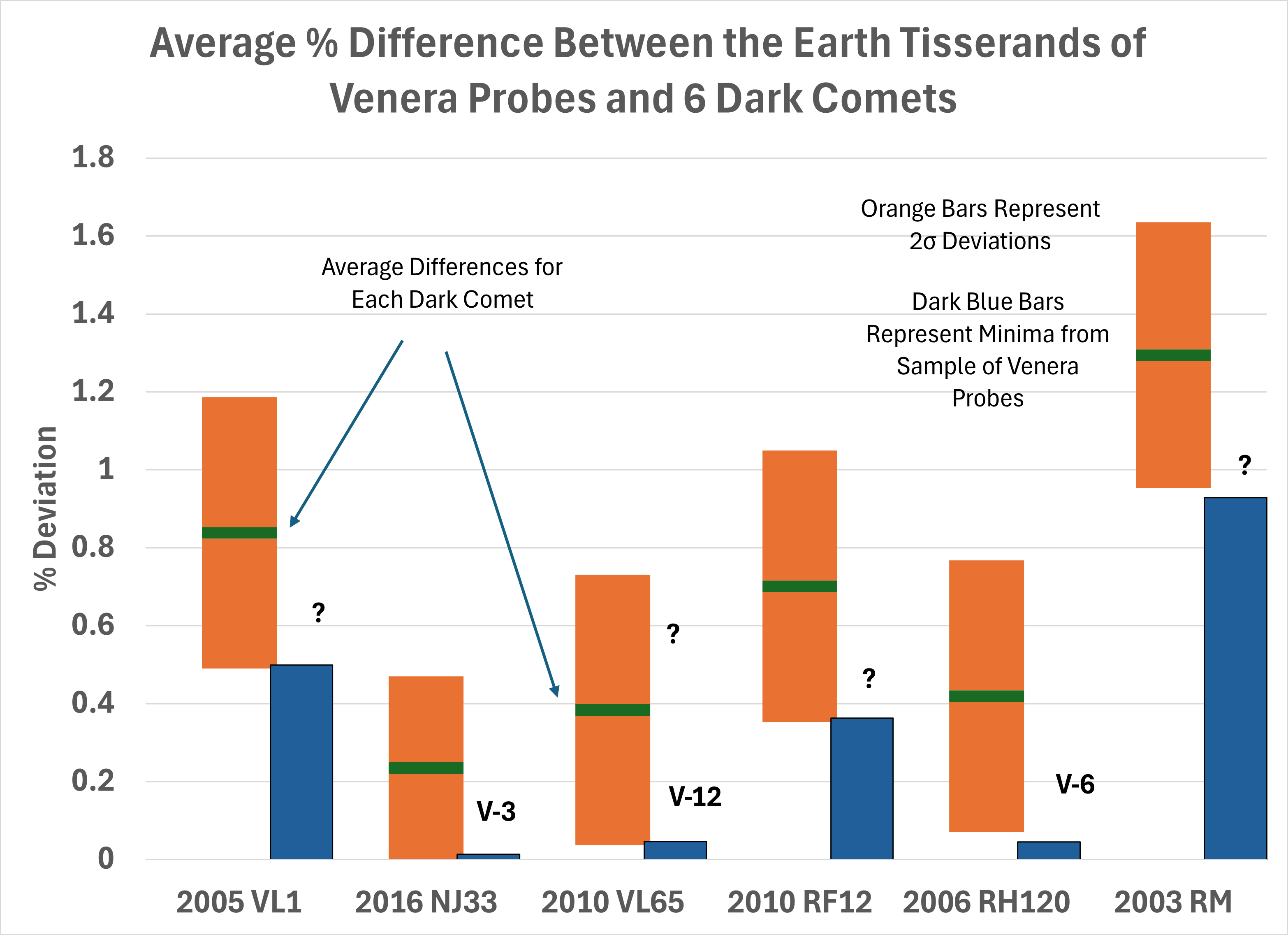}
\caption{Barchart of average deviation of Earth Tisserands for the Venus probes with respect to each of the Dark Comets listed in Table \ref{TAB3}. The orange error bars indicate the range of these deviations (2$\sigma$)}.
\label{fig:Bar3}
\end{figure}

We now consider the Earth Tisserands computed for each of the Russian Venus probes listed in Table \ref{table:TAB1} and compare them against each of the Dark Comets listed in Table \ref{TAB3}. (Note \textit{Venera-3} crashed into the planet Venus on 1966 March 1$^{st}$ \citep{1985JBIS...38...74C}, but we are considering the Blok-L upper stage in this analysis.) To this end refer to Figure \ref{fig:Bar3}.\\

The general level of disagreement ($2\sigma$) in Tisserand between the probes and the Dark Comets is $\sim{1 \%}$. Observe that the value of the mean for the best fit Exponential distribution for Figure \ref{fig:Hist2} is $\sim{5.81\%}$, equivalent to a $2\sigma$ of $11.62 \%$, much larger than the general disagreement of Tisserands with the probes.\\

Now we refer to Figure \ref{fig:Bar3} and we find that for $2005\ VL_1$, all of the \textit{Venera} probes are displaced significantly from Tisserand of $2005\ VL_1$, and furthermore, so is the minimum deviation of the probes. Thus the question mark indicates that we cannot firmly attribute ANY of the \textit{Venera} probes to this particular object. This is also the situation for $2010\ RF_{12}$ \& $2003\ RM$.
 For the remaining 3 Dark Comets, $2016\ NJ_{33}$, $2010\ VL_{65}$ and $2006\ RH_{120}$, the \textit{Veneras} 3, 12, \& 6 respectively are the most likely candidates, with deviations of 0.014\%, 0.05\% and 0.05\% . All these, but especially the last 2, represent discrepancies in the Tisserands of around the alteration in its value expected from an Earth encounter, as determined empirically in Section \ref{sec2sub5} and shown in Figures \ref{fig:TISS1} and \ref{fig:TISS2}.\\

\begin{figure}[hbt!]
\centering
\includegraphics[width=0.8\textwidth]{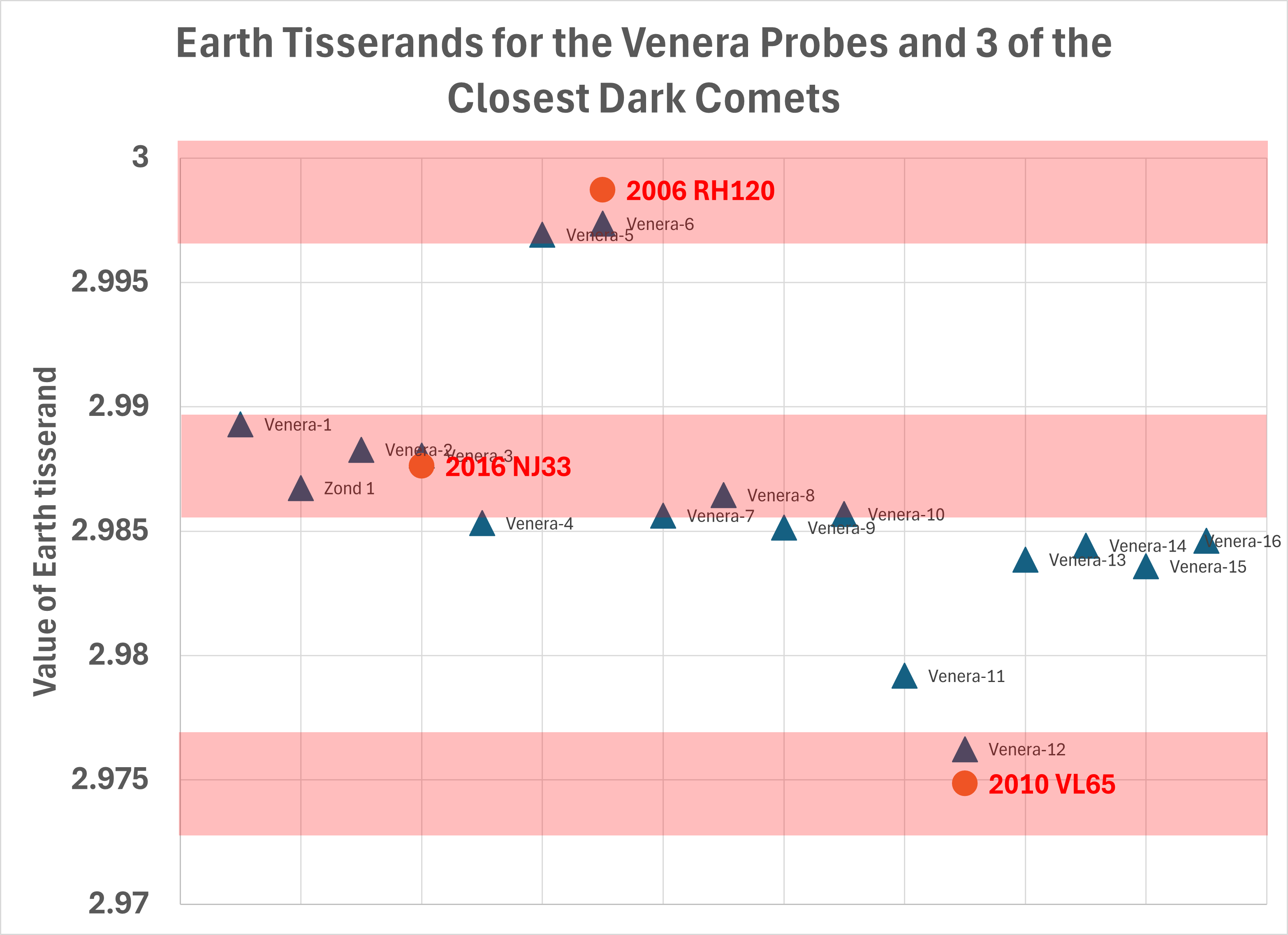}
\caption{Scatter of Earth Tisserand parameters for all the \textit{Venera} probes and also the 3 Dark Comets to which 3 \textit{Veneras} are very close. The $2\sigma$ possible alterations of the Tisserands (equivalent to $\sim{0.056}$\%), due to the passage of time, as established in Section \ref{sec2sub5}, are indicated by the shaded regions either side of each Dark Comet.}
\label{fig:Scat}
\end{figure}

\begin{table}[]
\renewcommand{\arraystretch}{1.6}
\centering
\begin{tabular}{|cccc|}
\hline
\textbf{\begin{tabular}[c]{@{}c@{}}Dark \\ Comet\end{tabular}} &
  \textbf{\begin{tabular}[c]{@{}c@{}}Venus \\ Probe\end{tabular}} &
  \textbf{\begin{tabular}[c]{@{}c@{}}Discrepancy in\\ Earth Tisserand\end{tabular}} &
  \textbf{\begin{tabular}[c]{@{}c@{}}\% of random \\ NEOs as close\end{tabular}} \\ \hline
\textbf{$2016\ NJ_{33}$}   & \textit{Venera-3} & 0.014\% & 0.23 \\ \hline
\textbf{$2006\ RH_{120}$} & \textit{Venera-6}  & 0.05\% & 0.75  \\ \hline
\textbf{$2010\ VL_{65}$} & \textit{Venera-12}  & 0.05\% & 0.75   \\ \hline
\end{tabular}
\caption{Possible Attributions as a result of analysis conducted in Section \ref{sec2sub6}.\label{TAB4}}
\end{table}

\section{Discussion}
\label{sec3}

To summarise, the results of Section \ref{sec2sub6} are provided in Table \ref{TAB4}.\\

To clarify what is happening with the \textit{Veneras} and the Dark Comets, refer now to Figure \ref{fig:Scat}, which is a scatter plot showing the Earth Tisserands of all the \textit{Veneras} and also the Dark Comets to which 3 are closely associated as summarised in Table \ref{TAB4}. The red bands indicate the 2$\sigma$ possible change for any \textit{Venera} probe over the course of a 50 year timescale (refer Section \ref{sec2sub5}).\\

We find that the association of \textit{Venera-3} with Dark Comet $2016\ NJ_{33}$ is not entirely unusual in that there are at least 6 other \textit{Veneras} which could be similarly ascribed. For the association of \textit{Venera-6} to $2016\ RH_{120}$
, this is certainly within the realms of possibility, since \textit{Venera-6} (and \textit{Venera-5} for that matter) are just inside the $2\sigma$ limits determined in Section \ref{sec2sub5}. Finally we can conclude that the association of $2010\ VL_{65}$ with \textit{Venera-12} is quite significant, and it is unlikely ANY other of the $\textit{Veneras}$ could be this object.\\

Let us take each of the 3 \textit{Venera} probes shown in Table \ref{TAB4}, and trace back their respective histories, in order to establish some further evidence or not for the likelihood of the attributions made in this study.\\

\textit{Venera-12} was launched on 1978 Sep 14$^{th}$ and arrived at Venus on 1978 Dec 21$^{st}$. According to \cite{1985JBIS...38...74C}, although a lander was successfully deployed, nevertheless the main spacecraft bus flew past Venus at a pericythe of $35,000$ \si{km} and continued into a heliocentric orbit. Furthermore, \cite{2023LPICo2851.2036S} give the Dark Comet of $2010\ VL_{65}$ a radius, $R_{NUC}$, of $\sim{3}$ \si{m} (Table 1 of \cite{2023LPICo2851.2036S}), corresponding to a diameter $\sim{6}$ \si{m}. However in practice this makes various assumptions about the nature of the object, including an average albedo, and assumes a general spherical shape. \cite{2023LPICo2851.2036S} also make a comment that all the Dark Comets are small and of size $\sim{3\ to\ 15}$ \si{m} radius which fits with the \textit{Venera-12} Blok-L. \\

\textit{Venera-6} launched on 1969 Jan 10$^{th}$, with arrival on 1969 May 17$^{th}$, having separated into a lander section and bus at a distance of $25,000$ \si{km} from Venus. The mission was a success it seems and the bus continued on into a heliocentric orbit. The value of $R_{NUC}$ for $2016\ RH_{120}$ in \cite{2023LPICo2851.2036S} is in the range 2-7 \si{m}, though the caveat mentioned for $2010\ VL_{65}$ also applies here.\\

Finally \textit{Venera-3}. Launched on 1965 Nov 16$^{th}$, and impacted Venus on 1966 Mar 1$^{st}$ after a course correction on 1965 Dec 26$^{th}$. However as already articulated, this does not preclude  $2016\ NJ_{33}$ from being associated with \textit{Venera-3} in some way, since the Blok-L for this mission may be an alternative candidate. The value of $R_{NUC}$ for $2016\ NJ_{33}$ is $\sim{16}$ \si{m}. Now let us examine the three Dark Comets in more detail.\\

 The Dark Comet $2006\ RH_{120}$ was initially given the Catalina Sky-Survey designation 6R10DB9. There were some initial suggestions that the object could be artificial, and the presence of Solar Radiation Pressure was shown to be perturbing its path. Nonetheless because this is a relatively small body, it was eventually categorized as consistent with a rocky object, and received its official designation in 2008. The presence of Solar Radiation Pressure on $2006\ RH_{120}$ can be deduced by examining the non-gravitational $A_1$ term (radial) in Table 1 of \cite{2023LPICo2851.2036S} and has a significance of 18$\sigma$.   \\ 

In the paper \cite{2023LPICo2851.2036S}, it is clear that the object $2016\  NJ_{33}$ also has an especially significant value of non-gravitational acceleration in the radial direction, $A_1$, to a confidence of 3$\sigma$. \\

It is interesting to discover that two of the Dark Comets attributable to \textit{Venera} probes in Table \ref{TAB4}, happen also to have the most significant non-gravitational radial accelerations of those Dark Comets provided in Table 1 of \cite{2023LPICo2851.2036S}.

\section{Conclusion}
\label{sec4}
In this paper, we found that the nature of the transfer orbits of the Russian \textit{Venera} probes led to a capability for them to return to Earth, although it is doubtful that the mission planners had originally intended them to do so. We provided the orbital parameters of the \textit{Venera} probes from launch and arrival dates. We found that all the Dark Comets have an Earth Tisserand remarkably close to 3 with a  probability of 0.0007 for a randomly selected set of 6 NEOs. We determined that the invariance of the Earth Tisserand in the presence of an encounter with Earth, for a \textit{Venera} type orbit is maintained generally around $1\sigma = 0.028\%$ over a period of 50 years. We discovered that the \textit{Venera} probes have similar values of Earth Tisserand to those of the Dark Comets. We deduced the most likely association of a Dark Comet with a \textit{Venera} mission would be the Blok-L upper stage. We found that 3 Dark Comets have Earth Tisserands close to those of 3 different \textit{Venera} missions. We established that two of these Dark Comets happen to have the most significant component of non-gravitational radial acceleration ($A_1$) of the 7 Dark Comets determined by \cite{2023LPICo2851.2036S}. In conclusion, the results suggest that the introduction of a new classification of celestial body, the Dark Comet, should be reconsidered until a fuller analysis can be conducted into the likelihood of them being interplanetary space junk.  
 \section{Acknowledgments}
 Brian Harvey, Matthew Williams \& Anatoly Zak.\\
\label{sec5}

\bibliography{Venera}{}
\bibliographystyle{aasjournal}
\end{document}